\documentclass[prb,preprintnumbers,amssymb]{revtex4}

\usepackage{amssymb}
\usepackage{amsmath}
\usepackage{graphicx}
\chardef\bslash=`\\ 

\begin{document}









\title{Anomalous Hall effect in ferromagnetic disordered metals}

\author{P. W\"olfle$^{1}$ and K.A. Muttalib$^{2}$}

\affiliation{$^{1}$Institut f\"ur Theorie der Kondensierten Materie, Universit\"at Karlsruhe, 76128 Karlsruhe, Germany\\
$^{2}$ Department of Physics, University of Florida, Gainesville, FL. 32611, U.S.A.}

\date{today}

\begin{abstract}

The anomalous Hall effect in disordered band ferromagnets is considered in
the framework of quantum transport theory. A microscopic model of electrons
in a random potential of identical impurities including spin-orbit coupling
is used. The Hall conductivity is calculated from the Kubo formula for both,
the skew scattering and the side-jump mechanisms. The recently discussed
Berry phase induced  Hall current is also evaluated within
the model. The effect of strong impurity scattering is analyzed and it is
found to affect the ratio of the non-diagonal (Hall) and diagonal components
of the conductivity as well as the relative importance of different
mechanisms.

\end{abstract}

\maketitle

\section{ Introduction}
\label{sect1}
\noindent
The Hall effect is known to occur in conductors subject to a magnetic field.
On a classical level it is explained by the Lorentz force acting on the
charge carriers in a magnetic field. However, an external (or internal)
magnetic field is not necessary for a Hall response to exist. It follows
from basic principles of statistical mechanics that the Hall effect can
appear whenever time reversal invariance is broken. Indeed it was recognized
in the 1950's that a Hall effect should exist in ferromagnetic metals even
in the absence of an external magnetic field, if the magnetic polarization
of the spin system is coupled to the orbital motion by spin-orbit coupling.
An anomalous Hall effect in ferromagnets has been observed in many systems
(for example see \cite{PRS,CF}).
The direct coupling of the moving charge carriers to the magnetic field
generated by the spins is much too small to explain the experimental
observations.
\vskip .2cm
\noindent
In a pioneering paper, Karplus and Luttinger \cite{KL} worked out a theory of this
effect, in which they pointed out the existence of an additional term in the
velocity operator proportional to the gradient of any electrical potential
acting on the carriers and to a term acting like a magnetic field (not the
dipolar field). The latter has been identified recently as a Berry phase
term \cite{SN,ON,JQM,H}. It generates a
 Hall current in equilibrium, and should exist even in the absence of any
impurity scattering. The Hall conductivity derived from this mechanism is
proportional to the Berry phase curvature averaged over all occupied
conduction band states. The precise dependence of this quantity on the
ferromagnetic polarization and on the spin-orbit interaction depends on the details of the system considered.
\vskip .2cm
\noindent
At about the same time, in 1955, Smit \cite{S} described a different mechanism
known as skew scattering. It is based on the fact that electrons in a plane
perpendicular to the magnetic polarization scatter from an impurity
potential in an asymmetric fashion, if they feel the polarization via the
spin-orbit interaction. This mechanism has been worked out in great detail.
It yields a Hall conductivity approximately proportional to the longitudinal
conductivity (which is governed by impurity scattering at low temperatures),
to the ferromagnetic polarization and to the spin-orbit coupling (assumed to
be weak). The relation of this to the earlier works was discussed in \cite{L}.
\vskip .2cm
\noindent
Yet another mechanism for the anomalous Hall effect was proposed by Berger
\cite{B}, the "side-jump" \ mechanism. It is based on the observation that the
trajectory of an electron scattering off an impurity is shifted sidewise by
the action of the spin-orbit coupling in the presence of a ferromagnetic (or
antiferromagnetic) spin polarization. The effect gives rise to a Hall
conductivity independent of the density of impurities, i.e. of the mean free
path. The characteristic length replacing the mean free path is the shift of
the trajectory, which may be estimated to be of the order of the lattice
spacing. Hence this contribution is small compared to the skew scattering
term , except in the case of short mean free path. Although this mechanism
is similar to the Berry phase mechanism, and may be shown to originate from
the extra term in the velocity operator, it involves the nonequilibrium quasiparticle distribution.
\vskip .2cm
\noindent
Quantum corrections to the anomalous Hall conductivity have not received
much attention so far. The weak localization correction is cut off by the
spin-orbit as well as the phase relaxation rates. Nonetheless, it has been
found to be of order unity in the disorder parameter $(k_{F}l)^{-1}$ ,
within the skew scattering mechanism \cite{LW,DCB}. In the case of the side-jump mechanism, weak localization corrections
have been found to be negligibly small \cite{DCB}. Interaction corrections
have been shown to be absent within the skew scattering model (neglecting
Hartree terms) in \cite{LW}, in the limit of weak impurity scattering.  For strong impurity scattering, however, a finite interaction correction appears \cite{MP}, in accordance with experimental observation \cite{MH}.
\vskip .2cm
\noindent
In this paper we will review these different mechanisms of the anomalous
Hall effect from a common perspective, such that their dependences on
parameters are displayed and their relative magnitudes are estimated. We
will limit our discussion to two-dimensional or quasi two-dimensional
disordered metallic band ferromagnets. As a convenient and not unrealistic
model of disorder we assume identical short range impurity potentials at
random positions, including spin-orbit interaction induced by the impurity
potential. The strengths of the impurity potential and of the spin-orbit
coupling will be left as free parameters. In particular we will be
interested in strong impurity scattering, which has not been considered
previously in this context within such a model, to our knowledge.

\vskip .2cm

\section{The model}
\label{sec2}

\noindent
We consider ferromagnetic metallic films with conduction electrons occupying
a spin-split band. Transport at low temperatures is governed by impurity
scattering. We will model the disorder by assuming identical impurities of
density $n_{imp}$ \ , at random positions $\mathbf{R}_{i}$ .
Electron-electron interaction effects will be neglected. Spin-orbit
interaction at the impurities will give rise to an anomalous Hall effect, as
pointed out in the early papers by Smit \cite{S}  and Luttinger \cite{L}. This
so-called skew scattering arises because the finite magnetization $\mathbf{M}
$ of \ the conduction electrons introduces a sense of rotation about the
direction of $\mathbf{M}$ . In addition there is a "side-jump " contribution
(\cite{B}) , caused by a sideways shift of the scattering wave packet due
to the spin orbit interaction. The spin-orbit interaction is a relativistic
effect and therefore rather small for transition metal atoms. As pointed out
in \cite{B}, the mixing of different d-orbitals provides a
renormalization of the coupling constant leading to an enhancement by a
factor of $10^{4}$ . We will take this effect into account by employing a
phenomenological coupling constant $g_{\sigma }$ of order unity.
\noindent
The single particle Hamiltonian of a conduction electron in a ferromagnetic
disordered metal , including spin-orbit interaction induced by the disorder
potential $V_{dis}(\mathbf{r)}$, is given in its simplest form by

\begin{equation}
\label{eq:1}
H_{1}=[-\frac{\nabla ^{2}}{2m}+V_{dis}(\mathbf{r)]\delta }_{\sigma \sigma
\prime }-M_{z}\tau _{\sigma \sigma \prime }^{z}-i(g_{\sigma }/4\pi n_{\sigma
})[\mathbf{\tau }_{\sigma \sigma \prime }\cdot (\nabla V_{dis}\times \nabla
)],
\end{equation}

\noindent
where\ $n_{\sigma }=(k_{F\sigma }^{2}/4\pi )$ and $k_{F\sigma}$ are  the density of
conduction electrons and the Fermi wave vector of spin $\sigma $ , $ \mathbf{\tau }_{\sigma \sigma
\prime }$  is the vector of Pauli matrices, and $g_{\sigma }$ is a
dimensionless spin-orbit coupling constant. The bare coupling constant is
given by $g_{\sigma }^{(0)}=(4\pi n_{\sigma })(2mc)^{-2}$ with $m$ the electron mass and $c$ the velocity of light . Note that $
g_{\sigma }$ is in general spin dependent. We use units such that  $M_{z}$
is half  the Zeeman energy splitting of
the conduction electron energies caused by
the ferromagnetic polarization $\mathbf{M}$. Here  $\mathbf{M}$ is assumed to be oriented along $z$, perpendicular to the layer . The disordered potential
will be modelled as  $V_{dis}(\mathbf{r})= \Sigma_{j}$
$V(\mathbf{r-R}_{j})$.  We will later average over the impurity positions
$\mathbf{R}_{j}$.
\vskip .2cm
\noindent
The matrix elements of  $H_{1}$ in the plane wave representation are
given by
\begin{equation}
\label{eq:2}
\langle \mathbf{k\prime \sigma \prime |} H_{1}| \mathbf{k\sigma}
\rangle = \Big(\frac{k^{2}}{2m}-M\sigma \Big)\ \mathbf{\delta}_{\mathbf{kk}\prime
}\mathbf{\delta }_{\sigma \sigma \prime }+\ \ \ \sum_{j}V(
\mathbf{k-k\prime} )\exp [i(\mathbf{k-k\prime} )\cdot \mathbf{R}_{j}]\{
{\mathbf{\delta}_{\sigma \sigma \prime }}\; - \; ig_{\sigma }
{\mathbf{\tau }_{\sigma \sigma \prime }}\cdot (\widehat{\mathbf{k}}
{\mathbf{\times\widehat{\mathbf{k}}\prime})\}}
\end{equation}
\vskip .2cm
\noindent
where $V(\mathbf{k-k\prime )}$\ is the Fourier transform of the single
impurity potential, and \ $\widehat{\mathbf{k}}=\mathbf{k/|k|}$ .
\vskip .2cm
\noindent
The many-body Hamiltonian is given in terms of electron creation and
annihilation operators $c_{\mathbf{k\sigma }}^{+},c_{\mathbf{k\sigma }}$ for
Bloch states $\mid\mathbf{k\sigma >}$ \ as
\begin{eqnarray}
\label{eq:3}
H &=& \ \sum_{\mathbf{k\sigma }}(\varepsilon _{\mathbf{k}
}-M_{z}\sigma )c_{\mathbf{k\sigma }}^{+}c_{\mathbf{k\sigma }}+\nonumber \\
&& +\sum_{\mathbf{k\sigma ,k\prime \sigma \prime }}
\sum_{j} V(\mathbf{k-k\prime )}\exp [i(\mathbf{k-k\prime )\cdot }
\ \mathbf{R}_{j}]\{\mathbf{\delta }_{\sigma \sigma \prime }\; - \;ig_{\sigma }
\mathbf{\tau }_{\sigma \sigma \prime }\cdot (\widehat{\mathbf{k}
}\mathbf{\times \widehat{\mathbf{k}}\prime )\}}c_{\mathbf{k\prime \sigma
\prime }}^{+}c_{\mathbf{k\sigma }},
\end{eqnarray}
\vskip .5cm
\section{ Normal and anomalous conductivity in
the limit of weak impurity scattering}
\label{sect3}

\vskip .5cm

\subsection{\bf Kubo formula and single particle Green's function}
\vskip .2cm
\noindent
The conductivity $\sigma _{\alpha \beta \text{ \ }}$ will be calculated from
the current response functions $L_{\alpha \beta }(\Omega _{m})$ by employing
the Kubo formula
\begin{equation}
\label{eq:4}
\sigma _{\alpha \beta \text{ \ }}=e^{2}\underset{\Omega _{m}\rightarrow 0}{
\lim }\frac{1}{\Omega _{m}}L_{\alpha \beta }(i\Omega _{m})\;\; , \;\;\;
L_{\alpha \beta }(i\Omega _{m})=\int_{0}^{\beta}d\tau e^{i\Omega
_{m}\tau }\langle T_{\tau }[j_{\alpha }(\tau )j_{\beta }(0)]\rangle
\end{equation}
\vskip .2cm
\noindent
where $j_{\alpha }(\tau )$ \ are the components of the current density in
the Heisenberg representation , \ $\Omega _{m} = 2\pi Tm$ \ are bosonic Matsubara
frequencies and the angular brackets denote the thermal average. We will
calculate the current correlation function within diagrammatic perturbation
theory in the impurity potential, averaging over the random impurity
positions. To simplify the calculations, we will drop the momentum
dependence of the single impurity potential, $V(\mathbf{k)=}V_{0}$ . The
momentum dependence of the spin-orbit term in the potential will be kept,of
course, as it is the source of the anomalous Hall effect.
\vskip .2cm
\noindent
The single particle Green's function $G_{\mathbf{k}\sigma }(i\omega _{n})$ \
is defined in terms of the self-energy $\Sigma _{\mathbf{k}\sigma }(i\omega
_{n})$ as
\begin{equation}
\label{eq:5}
G_{\mathbf{k}\sigma }(i\omega _{n})=[i\omega _{n}-\varepsilon _{\mathbf{k}
\sigma }-\Sigma _{\mathbf{k}\sigma }(i\omega _{n})]^{-1}
\end{equation}
\vskip .2cm
\noindent
Here $\omega _{n}=\pi T(2n+1)$   are fermionic Matsubara frequencies, and $\varepsilon _{\mathbf{k}\sigma
}=\varepsilon _{\mathbf{k}}-M_{z}\sigma $ \ are the Bloch energies. The
Bloch states are filled up to the Fermi energy $\varepsilon _{F\sigma }$ for
each subband.
\vskip .2cm
\noindent
In lowest order in the impurity potential, assuming all momenta to lie in
the x-y-plane and after averaging over the random positions of the
impurities we have
\begin{equation}
\label{eq:6}
\Sigma_{\mathbf{k}\sigma }(i\omega_{n})=n_{imp}V_{0}^{2}\sum_{
\mathbf{k}\prime }\{ 1 - ig_{\sigma }\mathbf{\tau}_{\sigma \sigma }^{z}\cdot (
\widehat{\mathbf{k}}\mathbf{\times \widehat{\mathbf{k}}\prime )}_{z}
\}^2 \;\; G_{\mathbf{k\prime }
\sigma }(i\omega_{n})\
\end{equation}
\vskip .2cm
\noindent
Since the Green's function
as a function of $\varepsilon _{\mathbf{k}\sigma }$ is strongly peaked at
the Fermi energy, one may separate the $\mathbf{k}\prime $-summation into an
integral over energy, taking the momenta, in particular that of $\Sigma _{
\mathbf{k}\sigma }(i\omega _{n})$ to be at the Fermi surface, and an
integral over the angle $\varphi \prime $ formed by $\mathbf{k}\prime $ and
the x-axis (note $\mathbf{\widehat{\mathbf{k}}\prime }=(\cos \varphi \prime
,\sin \varphi \prime )$. As a result, one finds
\begin{equation}
\label{eq:7}
\Sigma _{\mathbf{k}\sigma }(i\omega _{n})=-isign(\omega _{n})n_{imp}(\pi
N_{\sigma })^{-1}[w_{\sigma }+2u_{\sigma }]=-isign(\omega _{n})(2\tau
_{\sigma })^{-1}
\end{equation}
\vskip .2cm
\noindent
where $N_{\sigma }$ is the conduction electron density of states at the
Fermi level for spin subband $\sigma $ and $\tau_\sigma$ are the single particle relaxation times.  We have defined the dimensionless
coupling constants for potential scattering \ and spin-orbit scattering
\begin{equation}
\label{eq:8}
w_{\sigma }=(\pi N_{\sigma }V_{0})^{2} \,\,\,\, \mbox{{\rm and}}\;\;\;\; \ u_{\sigma }=(g_{\sigma
}/2)^{2}\ w_{\sigma } \ .
\end{equation}
\vskip .5cm

\subsection{\bf \protect\bigskip Longitudinal conductivity}
\vskip .2cm
\noindent
The diagonal elements of the conductivity tensor are given in lowest order
within our model by the bubble diagram. Assuming isotropic scattering
potential, there are no vertex corrections in lowest order. We have
\begin{equation}
\label{eq:9}
L_{\alpha \alpha }(i\Omega_{m})=T\sum_{\omega_{n}}\sum_{\mathbf{k},\sigma}
v_{\mathbf{k\sigma ,\alpha }}^{2}
G_{\mathbf{k}\sigma}(i\omega_{n})G_{\mathbf{k}\sigma }(i\omega_{n}-i\Omega_{m})
\end{equation}
\vskip .2cm
\noindent
where $v_{\mathbf{k\sigma ,\alpha }}=\partial \varepsilon _{\mathbf{k}\sigma
}/\partial k_{\alpha }=k_{\alpha }/m_{\sigma }$ \ is the particle velocity,
where $m_{\sigma }$ \ is the effective mass of quasiparticles with spin $
\sigma $ \ (we assume parabolic bands) . The integration over energy $
\varepsilon _{k}$ is only finite and equal to $2\pi \tau _{\sigma }N_{\sigma
}$ \ if the two poles of the G's are on opposite side of the real axis,
yielding the restriction on the frequency summation \ $0\leq \omega _{n}\leq
\Omega _{m}$ , \ and hence $\ (\Omega _{m}/2\pi T)$ (identical) terms. The
angular average over the velocity factors yields \ $\langle v_{\mathbf{
k\sigma ,\alpha }}^{2}\rangle =\frac{1}{2}v_{F\sigma }^{2}$ .\ As a result
one finds the sum of Drude conductivities for both spin orientations
\begin{equation}
\label{eq:10}
\sigma _{\alpha \alpha \text{ \ }}^{(0)}= e^2\;\;\sum_\sigma\;\;\frac{n_\sigma\tau_\sigma}{m_\sigma}\;\;\;= e^{2}\sum_{\sigma}D_{\sigma}^{(0)}
N_{\sigma }=e^{2}[4\pi n_{imp}V_{0}^{2}]^{-1}\sum_\sigma \,
v_{F\sigma }^{2}/(1+\frac{1}{2}g_{\sigma }^{2}) \ , \
\end{equation}
\vskip .2cm
\noindent
where \ \ $D_{\sigma }^{(0)}=\frac{1}{2}v_{F\sigma }^{2}\tau _{\sigma }$ \
is the diffusion constant. Thus, for weak scattering the conductivity is
found to be inversely proportional to (1) the density of impurities, (2) the
potential scattering cross section $V_{0}^{2}$ , and (3) the factor $\ (1+
\frac{1}{2}g_{\sigma }^{2})$ .\ Note that $D_{\sigma }^{(0)}N_{\sigma
}=\varepsilon _{F\sigma }\tau _{\sigma }/2\pi \gg 1$ , considering that in
the parabolic band approximation $N_{\sigma }=m_{\sigma }/2\pi $ .
\vskip .5cm
\subsection{\bf\protect\bigskip Anomalous Hall effect: Skew scattering contribution}

\vskip .2cm
\noindent
The lowest order diagram contributing to the Hall conductivity is a bubble
with three scattering processes at the same impurity, denoted by lines
running across (vertex correction needed to give finite angular averages),
shown in Fig. 1a,
\begin{eqnarray}
\label{eq:11}
L_{xy}(i\Omega_{m})&=&n_{imp}T\sum_{\omega_{n}}
\sum_{\mathbf{k},\mathbf{k\prime ,}\sigma}v_{\mathbf{k\sigma ,}x}
v_{\mathbf{k\prime \sigma ,}y}G_{\mathbf{k}\sigma }(i\omega_{n})G_{\mathbf{k}\sigma
}(i\omega_{n}-i\Omega_{m})\nonumber \\
&&\sum_{\mathbf{k}\prime\prime}\;\; V_{\mathbf{kk\prime \prime }}
G_{\mathbf{k\prime\prime }\sigma }
(i\omega_{n})V_{\mathbf{k\prime \prime k\prime }}G_{\mathbf{
k\prime }\sigma }(i\omega_{n})G_{\mathbf{k\prime }\sigma }(i\omega
_{n}-i\Omega_{m})V_{\mathbf{k\prime k}}
\end{eqnarray}
\vskip .2cm
\noindent
and a  contribution with upper and lower line interchanged.$\ $Here
we use$\ \ \ V_{\mathbf{k\sigma ,k\prime \sigma \prime }}=
V_{0}\{1 - ig_{
\sigma }\mathbf{\tau }_{\sigma \sigma }^{z}\cdot (\widehat{
\mathbf{k}}\mathbf{\times \widehat{\mathbf{k}}\prime )}_{z}\mathbf{\}}$.
\vskip .2cm
\noindent

\begin{figure}
\begin{center}
\includegraphics[width=0.36\textheight]{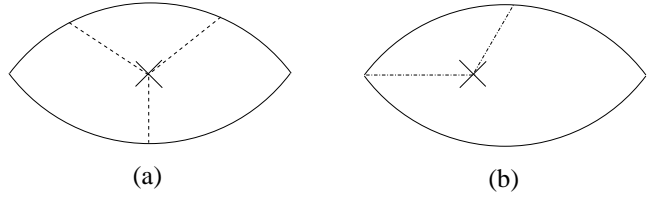}
\caption{\label{fig1} lowest order diagrams for (a) skew
scattering and (b) side-jump contributions. As described in the
text, there are two diagrams of type (a) and four of type (b). }
\end{center}
\end{figure}

Again the energy integrations on $\varepsilon _{k},\varepsilon _{k\prime }$
provide the restriction $0\leq \omega _{n}\leq \Omega _{m}$ \ and yield a
factor $(2\pi \tau _{\sigma }N_{\sigma })^{2}$ \ , while the energy
integration on the single $G_{\mathbf{k\prime \prime }\sigma }(i\omega _{n})$
\ gives $-isign(\omega _{n})\pi N_{\sigma }$ and the two angular
integrations are finite only for the cross terms in the product, using $
\langle \widehat{k}_{\alpha }\widehat{k}_{\beta }\rangle =\frac{1}{2}\delta
_{\alpha \beta }$.  The skew scattering
contribution to the Hall conductivity is then given by
\begin{equation}
\label{eq:12}
\sigma_{xy\text{ \ }}^{ss(0)}=\;e^{2}\sum_{\sigma }
\mathbf{\tau}_{\sigma \sigma }^{z}D_{\sigma }^{(0)}N_{\sigma }
\sqrt{w_{\sigma }}g_{\sigma}/(1+\frac{1}{2}g_{\sigma }^{2})
=  \frac{e^2}{4} \sum_\sigma \tau_{\sigma\sigma}^z \Big(\frac{n_\sigma}
{n_{imp}}\Big) \frac{1}{\sqrt{w_\sigma}}\frac{g_\sigma}{(1 + g_{\sigma}^2/2)^2}
\end{equation}
\vskip .2cm
\noindent
where in the last equality we used the definition of \ $D_{\sigma }^{(0)}$
and of \ $N_{\sigma }$ in the parabolic band approximation.
It is interesting to
note that the Hall conductivity is proportional to the spin-orbit coupling
for small \ $g_{\sigma }$ , while\ for large spin-orbit coupling, $g_{\sigma
}\gg 1$ , it is seen to decrease as $1/g_{\sigma }^{3}$ . This regime may be
reached only for sufficiently weak scattering, such that the conditions for
the validity of the above weak coupling calculation , \ $w_{\sigma
},u_{\sigma }\ll 1$ , are satisfied. In the absence of spin splitting , i.e.
without spontaneous ferromagnetic polarization , the factor \ $\mathbf{\tau
\ }_{\sigma \sigma }^{z}$ \ causes $\sigma _{xy\text{ \ }}^{ss(0)}$ \ to
vanish. In the limit of weak polarization $\mathbf{M}$ (along the z-axis,
perpendicular to the ferromagnetic layer), $\sigma _{xy\text{ \ }}^{ss(0)}$
\ is proportional to $M_{z}$ . \ The sign of $\sigma _{xy\text{ \ }}^{ss(0)}$
\ is given by the sign of $(-M_{z})$ , provided $D_{\sigma }^{(0)},N_{\sigma
},g_{\sigma }$ are increasing functions of the chemical potential, which
would be the case for a parabolic energy spectrum. The ratio of anomalous
Hall conductivity and longitudinal conductivity is small, proportional to
(1) the ferromagnetic Zeeman energy in units of the Fermi energy $\
(M_{z}/\varepsilon _{F})$ \ (2) the spin-orbit coupling constant \ $%
g_{\sigma }$ (for small coupling) \ (3) \ the dimensionless impurity
potential $(V_{0}N_{0})$ \ , where $N_{0}$ is the spin averaged density of
states at the Fermi energy:
\begin{equation}
\label{eq:13}
\sigma _{xy\text{ \ }}^{ss(0)}/\sigma _{\alpha \alpha \text{ \ }
}^{(0)}\simeq  \Big(\frac{M_{z}}{\epsilon_F}\Big)\;\;V_{0}N_0\;\;
\langle g_0 \rangle_\sigma
\end{equation}
\vskip .5cm
\subsection{\bf\protect\bigskip  Anomalous Hall effect: Side-jump contribution}
\vskip .2cm
\noindent
This effect may be calculated in a straightforward way \cite{CWDB,CB} by observing that the side-jump leads to an additional
term in the particle velocity due to the spin-orbit interaction. Indeed, the
quantum mechanical velocity obtained from the Heisenberg equation of motion
for the position operator has two terms,
\vskip .2cm
\begin{equation}
\label{eq:14}
\mathbf{v=}\frac{d}{dt}\mathbf{r=-}i\mathbf{[r,}H_{1}]=\frac{\mathbf{p}}{m}
+(g_{\sigma }/4\pi n_{\sigma })(\mathbf{\tau }_{\sigma \sigma }\mathbf{
\times \nabla }V_{dis})\ .
\end{equation}
\vskip .2cm
The matrix elements of $\mathbf{v}$ are given by
\begin{equation}
\label{eq:15}
\langle \mathbf{k\prime \sigma \prime |v}\, |\, \mathbf{k\sigma \rangle} =
\frac{\mathbf{k}}{m}\mathbf{\delta }_{\mathbf{kk}\prime }\mathbf{\delta }
_{\sigma \sigma \prime }-i(g_{\sigma }/4\pi n_{\sigma })\sum V(\mathbf{
k-k\prime )}{} e^{i(\mathbf{k-k\prime} )\cdot \mathbf{R}_{j}}{}
\{\mathbf{\tau}_{\sigma \sigma \prime }\times (\mathbf{k-\mathbf{k}\prime} )\}
\end{equation}
\vskip .2cm
\noindent
In lowest order in the impurity scattering, the side-jump contribution to
the Hall conductivity $\sigma _{xy}$ is calculated from four diagrams
with an impurity correlator line ended at one of the current vertices
and the upper or lower line, respectively (see Fig. 1b). Since they give identical contributions, we have to
consider only the first one. In 2d systems confined to the x-y plane, and
assuming the magnetization oriented in the z-direction, and the external
frequency $\Omega _{m}$ to be positive,
\vskip .2cm
\begin{eqnarray}
\label{eq:16}
L_{xy}^{sj,a}&=& n_{imp}T\sum_{\omega_{n}}
\sum_{\mathbf{k,k\prime ,\sigma }}{\sum}G_{\mathbf{k\prime }\sigma }
(\omega_{n})G_{\mathbf{k}\sigma }(\omega_{n})G_{\mathbf{k}\sigma }(\omega_{n}-
\Omega_{m})\times\nonumber \\
&\times &(-ig_{\sigma}V_{0}^{2}/\varepsilon_{F\sigma })
[\mathbf{\tau_{\sigma \sigma \prime}}\times \frac{(\mathbf{k-k\prime})}{2m}]_{x}
\frac{k_{y}}{m}
\end{eqnarray}
\vskip .2cm
\noindent
Performing the energy integrations and the angular integrations as in the
above, using the definition of \ $\tau_{\sigma }$ and multiplying the above
by a factor of four to account for all the diagrams, it follows that the
Hall conductivity is obtained as
\begin{equation}
\label{eq:17}
\sigma _{xy}^{sj}= (e^{2}/2\pi )\sum_{\sigma }[g_{\sigma }/(1+\frac{1}{2}
g_{\sigma }^{2})] \ \mathbf{\tau }\textbf{\ }_{\sigma \sigma }^{z}
\end{equation}
\vskip .2cm
\noindent
Remarkably,  $\sigma _{xy}^{sj}$  is independent of the impurity
concentration. Since the effective mean free path characteristic of the
side-jump contribution is rather short, of order $g_{\sigma }/k_{F}$ , it
will be important only for dirty samples, when the skew scattering
contribution is also small.   A change in sign of the
skew scattering contribution has been found for high impurity concentration,
in the framework of the Coherent Potential Approximation \cite{CWDB}.
\vskip .5cm

\subsection{\bf\protect\bigskip Hall current in the clean limit}
\vskip .2cm
\noindent
The extra term in the velocity operator derived in Eq. (14) involves any potential $V(\mathbf r)$ acting on the conduction electrons.  In section 3.4 the potential considered was that due to impurities, $V_{dis}(\mathbf r)$.  In the pesence of an applied electric field $\mathbf{E}$, putting $\mathbf{\nabla} V$ in
Eq. (14) equal to $-e \mathbf{E}$, the extra term in the velocity is directed orthogonal to $\mathbf{E}$ and to the magnetization $\mathbf{M}$, and hence may give rise to a Hall current.  Since the velocity operator in this case is itself proportional to the applied field, to obtain the linear response current it is sufficient to average it with the equilibrium statistical operator resulting in
\begin{equation}
j_x{} = {}  e^2 \;\sum_{\mathbf{k}\sigma} \;\frac{g_\sigma}{4\pi n_\sigma}\;\tau_{\sigma\sigma}^z E_{y}f_{k\sigma}
\label{17a}
\end{equation}
\noindent
where $f_{\mathbf{k}\sigma}$ is the Fermi function. With the help of $\sum_{\mathbf{k}}f_{\mathbf{k}\sigma} = n_\sigma$, one finds the contribution to the
Hall conductivity even in the absence of impurities (``clean limit''):
\begin{equation}
\sigma_{xy}^{c} = \frac{e^2}{4\pi}\;\;\sum_{\sigma}\;\;g_{\sigma}\;\;\tau_{\sigma\sigma}^z\;.
\label{17b}
\end{equation}
\noindent
This current flows in equilibrium and does not require the redistribution of particles from excited states into the equilibrium state.  It is therefore similar to a ballistic current.
\vskip .2cm
\noindent
It is a single particle current and it is therefore not stabilized or proteced by collective effects.  For that reason one may wonder how strongly this current decays as a consequence of inelastic or dephasing process.  This will be subject of future investigations.
\vskip .2cm
\noindent
Recently a  Hall current in a perfect crystal lattice of rather similar form has been
discussed \cite{SN,ON,JQM,H}  It arises due to a Berry phase acquired by electrons moving in the periodic potential of the crystal.  A Bloch electron moving in reciprocal space under the influence of the combined effect of spin-orbit interaction and a ferromagnetic polarization along a path $C$ acquires a Berry phase
\begin{equation}
\chi(\mathbf{k}) = - \int_{c}^{\mathbf{k}}\;\;d{\mathbf{k}\prime} \cdot \mathbf{X}({\mathbf{k}\prime})\;\;.
\label{17c}
\end{equation}
\vskip .2cm
\noindent
Here the Berry vector potential $\mathbf{X}(\mathbf{k})$ (of dimension length, and thus describing a shift of the Wannier coordinate of the Bloch states within the unit cell), is given by
\begin{equation}
\mathbf{X}(\mathbf{k}) = \int_{\rm cell}\;d^2r\;\;u_{n\mathbf{k}}^{\ast}
(\mathbf{r})\;i\mathbf{\nabla}_k\;\;u_{n \mathbf{k}}(\mathbf{r})\;\;.
\label{17d}
\end{equation}
\vskip .2cm
\noindent
Associated with $\mathbf{X}(\mathbf{k})$ is a ``magnetic field''
\begin{equation}
\mathbf{\Omega}(\mathbf{k})  = \mathbf{\nabla}_k \times \mathbf{X}(\mathbf{k})
\label{17e}
\end{equation}
\noindent
acting in $\mathbf{k}$-space.
\vskip .2cm
\noindent
The quasi-classical dynamics of Bloch electrons including the Berry phase may be derived from the Bloch Hamiltonian
\begin{equation}
H_{\mathbf{k}} = V (i\mathbf{\nabla}_k + \mathbf{X}_k) + \epsilon_{n\sigma}(\mathbf{k})\;\;,
\label{17f}
\end{equation}
\noindent
where $\epsilon_n(\mathbf{k})$ are the Bloch energies (including the spin-orbit interaction and the ferromagnetic polarization), and $V(\mathbf{r})$ is the applied external potential.  Putting $\mathbf{\nabla}V(\mathbf{r}) = - e \mathbf{E}$, the quasiclassical equations of motion derived from this effective Hamiltonian are
\begin{eqnarray}
\dot{\mathbf{k}} &=& e\mathbf{E} + e \dot{\mathbf{r}} \times \mathbf{B}\nonumber \\
\dot{\mathbf{r}} &=& \mathbf{\nabla}_{\mathbf{k}}\epsilon_{n\mathbf{k}} - e \mathbf{E} \times \mathbf{\Omega}
\label{17g}
\end{eqnarray}
\noindent
The additional term in the velocity $\dot{\mathbf{r}}$
leads to a Hall current
\begin{equation}
\mathbf{j}_H = - e^2n \;\langle \mathbf{\Omega}\rangle \times \mathbf{E}
\rightarrow \;\;\;\;\;\sigma_{xy}^B = e^2n \;\;\;\langle\Omega_z\rangle
\label{17h}
\end{equation}
\noindent
where $\langle \mathbf{\Omega} \rangle = n^{-1} \sum_{\mathbf{k}\sigma}\;\;\mathbf{\Omega}_\sigma (\mathbf{k})f(\epsilon_{\mathbf{k} \sigma})$ is the average of the Berry magnetic field over all occupied states in $\mathbf{k}$-space and $n = \Sigma_{\sigma}n_\sigma$.  This average is zero unless time reversal symemtry is broken, e.g. in a ferromagnet, and the spin polarization is coupled to the orbital motion.  This Hall current has been computed for ferromagnetic semiconductors \cite{JQM},  and found to be of the magnitude observed in such systems.
\vskip .2cm
\noindent
It would be instructive to compare the two types of  Hall currents discussed in this subsection within the same model.  The Berry phase mechanism is derived from the topological properties of Bloch states in a priodic potential, moving in $\mathbf{k}$-space under the influence of an external field.  The mechanism discussed first does not require the presence of a periodic potential, but derives solely from the extra term in the velocity operator due to
spin-orbit interaction.

\vskip .5cm
\section{Strong impurity scattering}
\label{sect4}
\vskip .2cm
\noindent
The scattering potential of a single impurity is not necessarily small
compared to the Fermi energy. If the product $N_{\sigma }V_{0}$ \ is of
order unity, or even larger than unity, or, in other words, if the coupling
constants $w_{\sigma },u_{\sigma }$ \ are not small compared to unity,
repeated multiple scattering off the same impurity becomes important.In
technical terms, the low order (Born) transition amplitudes we have used in
the above must be replaced by the full scattering amplitude. As may be
expected, this results in a substantial renormalization of the
conductivities.
\vskip .2cm
\subsection{\bf Single particle scattering amplitude}
\vskip .2cm
The repeated scattering of an electron off a single impurity may be
described in terms of the dimensionless scattering amplitude
\begin{equation}
\label{eq:18}
f_{{\mathbf{k}}\sigma ,{\mathbf{k}}\prime \sigma \prime }=\pi N_{\sigma }\delta
_{\sigma \sigma \prime }[V_{{\mathbf{k}}\sigma ,{\mathbf{k}}\prime \sigma \prime
}+VGV+VGVGV+....]
\end{equation}
\vskip .2cm
\noindent
where G is the single particle Green's function$.$V is the bare interaction
with one impurity at \ $\mathbf{R=0}$.  In the following we
assume short-ranged interaction and again drop the k-dependence of \ $
V({\mathbf{k-k}}^{\prime }).$ The dependence on energy is governed by the sharply \
peaked G's, so that we may neglect the dependence on the magnitude of
momentum and may put $|\mathbf{k|=}k_{F\sigma }$ , the Fermi momentum for
spin orientation $\sigma .$ Then the energy integration in intermediate
states may be done,
\vskip .2cm
\noindent
The remaining integrations on angle may be done using (in 2d and neglecting
the angular dependence of $V(\mathbf{k-k}^{\prime })$ ), $\ \langle
\widehat{\mathbf{k}}\rangle =0$ ,
$\langle \widehat{\mathbf{k}}^{2}\rangle =\frac{1}{2}$,  $\ \langle (
\widehat{\mathbf{k}}\times \widehat{\mathbf{k}}_{1})_{z}
(\widehat{\mathbf{k}}_{1}\times \widehat{\mathbf{k}}
\prime )_{z}\rangle _{k_{1}}=-\frac{1}{2}(\widehat{\mathbf{k}}\cdot
\widehat{\mathbf{k}}\prime)$ ,
$\langle (\widehat{\mathbf{k}}\cdot \widehat{\mathbf{k}}_{1})
(\widehat{\mathbf{k}}_{1}\times \widehat{\mathbf{k}}\prime )_{z}
\rangle _{\mathbf{k}_{1}}=\frac{1}{2}(\widehat{\mathbf{k}}\times
\widehat{\mathbf{k}}\prime )_{z}$,  where we defined $\langle
O\rangle =\int \frac{d\varphi }{2\pi }O$. Within these approximations the
potential scattering and spin-orbit scattering terms do not mix. Odd terms in V do not depend on \ $s_{\omega _{n}}$
whereas even terms do. The terms of perturbation theory may be summed up to
infinity as a geometric series, with the result
\begin{equation}
\label{eq:20}
f_{\mathbf{k}\sigma ,\mathbf{k}\prime \sigma }=
\frac{\widetilde{w}_{\sigma }}{\sqrt{w_{\sigma
}}} - i\tau _{\sigma \sigma }^{z}(\widehat{\mathbf{k}}\times
\widehat{\mathbf{k}}\prime )\frac{2
\widetilde{u}_{\sigma }}{\sqrt{u_{\sigma }}}-is_{\omega _{n}}[\widetilde{w}
_{\sigma }+2\widetilde{u}_{\sigma }(\widehat{\mathbf{k}}\cdot \widehat{\mathbf{k}}\prime )].
\end{equation}
\vskip .2cm
\noindent
Here we defined \ $\widetilde{w}_{\sigma }=w_{\sigma }/(1+w_{\sigma })$, \
and \ $\widetilde{u}_{\sigma }=u_{\sigma }/(1+u_{\sigma })$, \ where \ $
w_{\sigma }=(\pi N_{\sigma }V_{0})^{2}$ \ and \ $u_{\sigma }=(g_{\sigma
}/2)^{2}w_{\sigma }$ , \ and all quantities depend on the spin orientation $
\sigma $ (suppressed in the following, except in the final expressions
involving spin summation).
\vskip .5cm

\subsection{\bf Single particle relaxation rate}
\vskip .2cm
\noindent
The single particle relaxation rate $\tau _{\sigma }^{-1}$ \ is obtained
from the imaginary part of the self energy
\begin{equation}
\label{eq:21}
\frac{1}{\tau _{\sigma }}=2s_{\omega _{n}}\textrm{Im}\Sigma _{{\mathbf{k}}\sigma
}(\omega _{n})=2s_{\omega _{n}}n_{imp}\textrm{Im}f_{\mathbf{k}\sigma ,
\mathbf{k}\sigma }=2\frac{n_{imp}}{\pi N_{\sigma }}(\widetilde{w}_{\sigma }+
2\widetilde{u}_{\sigma })
\ ,
\end{equation}
\vskip .2cm
\noindent
where $n_{imp}$ \ is the density of impurities (number of impurities per
volume). One observes that $\frac{1}{\tau _{\sigma }}$ \ is proportional to
the Fermi energy, the average number of impurities per electron and the
dimensionless factor \ $(\widetilde{w}+2\widetilde{u})$ \ , expressing the
effective scattering strength per impurity.
\vskip .5cm

\subsection{\bf Particle-hole propagator}

\vskip .2cm

\noindent
The particle-hole propagator $\Gamma _{\mathbf{kk\prime }}(\mathbf{q}
;\epsilon _{n},\epsilon _{n}-\Omega _{m})$ \ is an important ingredient of
vertex corrections of any kind. Here $\mathbf{k+q/2},\mathbf{k-q/2}$ \ are
the initial , $\mathbf{k\prime +q/2},\mathbf{k\prime -q/2}$ the final
momenta and \ $\epsilon _{n},\epsilon _{n}-\Omega _{m}$ \ are the Matsubara
frequencies of the particle and the hole line, respectively. $\Gamma $ \
satisfies the following Bethe-Salpeter equation \ (we have defined
dimensionless quantities $\Gamma ,t$ \ by multiplying both with a factor $
(2\pi N_{\sigma }\tau _{\sigma })$)
\vskip .2cm
\begin{eqnarray}
\label{eq:22}
\Gamma _{\mathbf{kk\prime }}(\mathbf{q};i\epsilon _{n},i\epsilon
_{n}-i\Omega _{m})&=& t_{\mathbf{kk\prime }}(\mathbf{q};i\epsilon
_{n},i\epsilon _{n}-i\Omega _{m})+(2\pi N_{\sigma }\tau _{\sigma })^{-1}
\nonumber \\
&&\times \ \underset{\mathbf{k}_{1}}{\sum } t_{\mathbf{kk}
_{1}}(\mathbf{q};i\epsilon _{n},i\epsilon _{n}-i\Omega _{m})G_{\mathbf{k}
_{1}+\mathbf{q}/2,\sigma }(i\epsilon _{n})\nonumber \\
&&G_{\mathbf{k}_{1}-\mathbf{q}
/2,\sigma }(i\epsilon _{n}-i\Omega _{m})\Gamma _{\mathbf{k}_{1}\mathbf{
k\prime }}(\mathbf{q};i\epsilon _{n},i\epsilon _{n}-i\Omega _{m})
\end{eqnarray}
\vskip .2cm
\noindent
The kernel of this equation is the impurity averaged particle-hole
scattering amplitude (we consider only the case of equal spin of particle
and hole)
\vskip .2cm
\begin{equation}
\label{eq:23}
t_{\mathbf{kk\prime }}(\mathbf{q};i\epsilon _{n},i\epsilon _{n}-i\Omega
_{m})= n_{imp}f_{\mathbf{k}+\mathbf{q}/2,\sigma ;\mathbf{k}\prime +\mathbf{q}
/2,\sigma }(i\epsilon _{n})
f_{\mathbf{k}\prime -\mathbf{q}/2,\sigma ;\mathbf{
k}-\mathbf{q}/2,\sigma }(i\epsilon _{n}-i\Omega _{m})
\end{equation}
\vskip .2cm
\noindent
It is useful to represent the operator $\ t_{\mathbf{kk\prime }}(\mathbf{q}
=0)$ \ in terms of its eigenvalues $\lambda _{m}$ . Assuming isotropic band
structure, the eigenfunctions $\chi _{m}(\widehat{\mathbf{k}})=\exp
(im\varphi )$  are those of the angular momentum operator component $L_{z}$.
The operator $ t_{\mathbf{kk}\prime }^{+-}(\mathbf{q}=0)$  may be represented as
\vskip .2cm
\noindent
\begin{equation}
\label{eq:31}
 t_{\mathbf{kk}\prime }^{+-}(\mathbf{q}=0)  =  {\sum_m }
\lambda _{m}\chi _{m}(\widehat{\mathbf{k}})\chi _{m}^{\ast }(\widehat{
\mathbf{k}}\prime ) \;\;\;{\textrm {and}}\;\;\;\;
 t_{\mathbf{kk}\prime }^{-+}(\mathbf{q}=0)
=[ t_{\mathbf{kk}\prime }^{+-}(\mathbf{q}=0)]^{\ast }
\end{equation}
\vskip .2cm
\noindent
The eigenvalues for $s_{p}= \rm{sign} (\epsilon_n) = 1,$  $s_{h}=
\rm{sign} (\epsilon_n - \Omega_m)  = -1$  are given by
\begin{equation}
\lambda _{0}=1\;\;\;,\;\;\;\;\;\;\;
\lambda _{1}=2\widetilde{w}\widetilde{u}(\widetilde{w}+2\widetilde{u}
)^{-1}(1+ is_{p}\frac{1}{\sqrt{u}}\tau_{\sigma \sigma }^{z})
\end{equation}
\vskip .2cm
\begin{equation}
\label{eq:30}
\lambda_{2}=\frac{\widetilde{u}^{2}}{u}(\widetilde{w}+2\widetilde{u}
)^{-1}(u-1 + 2is_{p}\sqrt{u}\tau_{\sigma \sigma }^{z})\;\;\;
{\textrm and}\;\; \; \lambda_{-m}=\lambda_{m}^{\ast}.
\end{equation}
\vskip .2cm
\noindent
The energy integral over the product of Green's functions in the integral
equation for $\Gamma _{\mathbf{kk\prime }}$  may be done first, after
expanding the G's in  $\Omega _{m}$  and  $q$,
where  $\mathbf{q\cdot v}_{k}=qv_{F}(\widehat{\mathbf{q}}\cdot \widehat{
\mathbf{k}})$ . Expanding $\Gamma _{\mathbf{kk\prime }}$ and  $ t_{
\mathbf{kk\prime }}$  in terms of eigenfunctions $\chi _{m}(\widehat{\mathbf{
k}})$ , $\Gamma _{\mathbf{kk\prime }}=$\ $\sum_{m}\Gamma _{m
\mathbf{k}\prime }\chi _{m}(\widehat{\mathbf{k}})$,  one obtains
\vskip .2cm
\noindent
\begin{eqnarray}
\label{eq:36}
\Gamma_{m\mathbf{k}\prime }=\lambda _{m}\chi _{m}(\widehat{\mathbf{k}}
\prime )&+&\lambda _{m}\Big\{\Big[1-\tau (|\Omega _{m}|+D_{0}q^{2})\Big]-\frac{i}{2}
v_{F}q\tau \Big[ \Gamma _{m-1\mathbf{k}\prime }\chi _{1}^{\ast }(\widehat{
\mathbf{q}})+\nonumber \\
& +&\Gamma _{m+1\mathbf{k}
\prime }\chi _{1}(\widehat{\mathbf{q}})\Big]-\frac{1}{4}(v_{F}q\tau )^{2}\Big[\Gamma
_{m-2\mathbf{k}\prime }\chi _{2}^{\ast }(\widehat{\mathbf{q}})+\Gamma _{m+2
\mathbf{k}\prime }\chi _{2}(\widehat{\mathbf{q}})\Big]\Big\}
\end{eqnarray}
\vskip .2cm
\noindent
The case $m=0$ \ needs special consideration, because particle number
conservation causes $\Gamma _{0\mathbf{k}\prime }$ to have a pole in the
limit $\Omega _{m},q\rightarrow 0$ , here expressed by $\lambda _{0}=1$ . \
\vskip .2cm
\noindent
The complete particle-hole propagator in the regime $v_{F}q\tau <1$ is given
by

\begin{equation}
\label{eq:40}
\Gamma _{\mathbf{kk}\prime }=\frac{1/\tau }{|\Omega _{m}|+Dq^{2}}\gamma _{
\mathbf{k}}\widetilde{\gamma }_{\mathbf{k}\prime }+\ \sum_{m\neq 0}\widetilde{\lambda }_{m}\chi _{m}(\widehat{\mathbf{k}})\chi _{m}^{\ast }(
\widehat{\mathbf{k}}\prime )\; \;\;;\;\;\;\; \widetilde{\lambda }_{m}=
\frac{\lambda_{m}}{1-\lambda_{m}}
\end{equation}

\begin{equation}
\label{eq:41}
\gamma _{\mathbf{k}}=1-\frac{i}{2}v_{F}q\tau \underset{m=\pm 1}{\sum }
\widetilde{\lambda }_{m}\chi _{m}(\widehat{\mathbf{k}})\chi _{m}^{\ast }(
\widehat{\mathbf{q}}\prime )=1-i\tau \underset{m=\pm 1}{\sum }\widetilde{
\lambda }_{m}\chi _{m}(\widehat{\mathbf{k}})\langle \mathbf{q\cdot v}
_{k\prime }\chi _{m}^{\ast }(\widehat{\mathbf{k}}\prime )\rangle _{\mathbf{
k\prime }}
\end{equation}

\begin{equation}
\label{eq:42}
\widetilde{\gamma }_{\mathbf{k}}=1-i\tau \underset{m=\pm 1}{\sum }
\widetilde{\lambda }_{m}\chi _{m}^{\ast }(\widehat{\mathbf{k}})\langle
\mathbf{q\cdot v}_{k\prime }\chi _{m}(\widehat{\mathbf{k}}\prime )\rangle _{
\mathbf{k\prime }}
\end{equation}
\vskip .2cm
\noindent
where the renormalized diffusion constant is defined as
\begin{equation}
\label{eq:39}
D_{\sigma }=D_{\sigma }^{(0)}\frac{1-\lambda _{1}^{\prime }}{|1-\lambda_{1}|^{2}} ,
\;\;\;\;{\textrm{ where}} \;\;\;\;\;
\lambda _{1}^{\prime }=\textrm{Re}\;\;\;\lambda _{1}  .
\end{equation}\noindent
\vskip .2cm
\noindent
The vertex corrections of the  current
vertices  $j_{\mathbf{k}\alpha }$  and $\widetilde{j}_{\mathbf{k}\alpha }$
 (for the incoming and outgoing current) are obtained by
\begin{eqnarray}
\label{eq:43}
j_{\mathbf{k\protect\sigma ,}\protect
\alpha }(q) &=& \mathbf{v}_{\mathbf{k}\protect\alpha }+\langle \mathbf{v}_{
\mathbf{k}\prime \protect\alpha }\Gamma_{\mathbf{k}\prime \mathbf{k}
}\rangle _{\mathbf{k}\prime }
\nonumber \\
\widetilde{j}_{\mathbf{k\sigma ,}\alpha }(q)
&=& \mathbf{v}_{\mathbf{k}\alpha}+\langle \mathbf{v}_{\mathbf{k}\prime \alpha }
\Gamma_{\mathbf{kk}\prime}\rangle_{\mathbf{k}\prime}
\end{eqnarray}
\vskip .2cm
\noindent
Note that $\widetilde{j}_{\mathbf{k}\alpha }\neq ($\ $j_{\mathbf{k}\alpha
})^{\ast }$ \ , \ as the eigenvalues $\widetilde{\lambda }_{m}$ are complex
valued, in general.
\vskip .5cm

\subsection{\bf  Conductivity tensor: Skew scattering}
\vskip .2cm
\noindent
Multiple scattering off the same impurity generates momentum dependence of
the effective scattering amplitude  and gives rise to vertex corrections
to the current vertex as calculated above. The current-current correlator is
now given by
\begin{equation}
\label{eq:44}
L_{\alpha \beta }(i\Omega _{m})=T\sum_{\omega _{n}}\sum_{
\mathbf{k},\sigma }v_{\mathbf{k\sigma ,\alpha }}\widetilde{j}_{\mathbf{k}
\sigma ,\beta }(q=0)G_{\mathbf{k}\sigma }(i\omega _{n})G_{\mathbf{k}\sigma
}(i\omega _{n}-i\Omega _{m})
\end{equation}
\vskip .2cm
\noindent
Using the explicit expressions for the components of the current vertex
\begin{equation}
\label{eq:45}
\widetilde{j}_{\mathbf{k\sigma ,}x}(q=0)=v_{F\sigma }[(1+\widetilde{\lambda
}_{1}^{\prime })\widehat{k}_{x}-\widetilde{\lambda }_{1}^{\prime\prime
}\widehat{k}_{y}]\;;\;\;\;
\widetilde{j}_{\mathbf{k\sigma ,}y}(q=0)=v_{F\sigma }[(1+\widetilde{\lambda
}_{1}^{\prime })\widehat{k}_{y}+\widetilde{\lambda }_{1}^{\prime\prime
}\widehat{k}_{x}]
\end{equation}
\vskip .2cm
\noindent
where the terms involving the diffusion pole drop out (they involve an angular average $\langle q_x\;q_y \rangle = 0$),
one finds for the conductivity tensor
\begin{equation}
\label{eq:47}
\sigma _{\alpha \beta \text{ \ }}^{ss}=e^{2}\sum_{\sigma }D_{\sigma
}^{(0)}N_{\sigma }
\begin{pmatrix}
1+\widetilde{\lambda }_{\sigma }^{\prime } &
 \widetilde{\lambda }_{\sigma
}^{\prime\prime } \\
-\widetilde{\lambda }_{\sigma }^{\prime\prime } & 1+\widetilde{\lambda
}_{\sigma }^{\prime }
\end{pmatrix}
\end{equation}
\vskip .2cm
\noindent
where we have defined $\widetilde{\lambda }_{\sigma}^{\prime }=\textrm{Re}
\widetilde{\lambda }_{1}^{\sigma}$
and
$\widetilde{\lambda }_{\sigma}^{\prime\prime}=\textrm{Im}\widetilde{
\lambda }_{1}^{\sigma }$.
We recall that $\widetilde{\lambda}_\sigma = \lambda_{1\sigma}/
(1-\lambda_{1\sigma})$ and $\lambda_{1\sigma}^\prime = 2\widetilde{w}\widetilde{u}(\widetilde{w} + 2\widetilde{u})^{-1}$, $\lambda_{1\sigma}^{\prime\prime} =  \lambda_{1\sigma}^\prime \frac{1}{\sqrt{u}}\tau_{\sigma\sigma}^z$.
Defining the tensor of diffusion coefficients
$D_{\alpha \beta }^{\sigma }$ as
\begin{equation}
\label{eq:48} D_{\alpha \alpha }^{\sigma
}=\frac{1}{2}v_{F\sigma}^{2}\tau _{\sigma }^{tr} \;\;, \;\;\;
D_{xy}^{\sigma }=D_{\alpha \alpha }^{\sigma } [\widetilde{\lambda
}^{\prime\prime}/(1+ \widetilde{\lambda }^{\prime
})]=-D_{yx}^{\sigma }
\end{equation}
\vskip .2cm
\noindent
where $\tau _{\sigma }^{tr}=\tau _{\sigma }(1+\widetilde{\lambda }^{\prime
}) $ \ is the momentum relaxation time, we may write
\vskip .2cm
\begin{equation}
\label{eq:50}
\sigma _{\alpha \beta }^{ss}=\sum_{\mathbf{\sigma }} N_{\sigma
}D_{\alpha \beta }^{\sigma } \ .
\end{equation}
\vskip .5cm
\subsection{\bf Side-jump mechanism}
\vskip .2cm \noindent In the strong scattering regime the bare
impurity scattering potential needs to be replaced by the
scattering amplitude. In addition the vertex corrections to the
current density operator have to be applied.  There are two new
diagrams involving the scattering line denoting the spin-orbit
term in the velocity operator, Eq. (14), and ending at one of the
current vertices, framed by two (instead of only one) scattering
amplitude lines (see Fig. 2).

\begin{figure}
\begin{center}
\includegraphics[width=0.36\textheight]{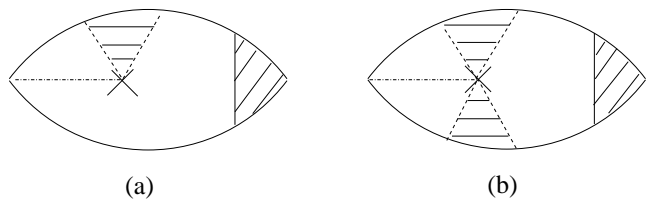}
\caption{\label{fig1} Two types of diagrams for the side-jump
contribution in the strong scattering regime. There are four
diagrams of type (a) and two diagrams of type (b). }
\end{center}
\end{figure}

The result of adding the six diagrams is
\begin{equation}
\sigma_{xy}^{sj} = \frac{e^2}{2\pi}\;\;\sum_\sigma\;\;\tau_{\sigma\sigma}^z
\frac{g_\sigma}{1 + \frac{1}{2}g_\sigma^2} \;\frac{1 +
\tilde\lambda^{\prime}_{1\sigma}}
{(1 + u_{\sigma})(1 + w_{\sigma})}
\label{eq:51}
\end{equation}
\vskip .5cm
\subsection{\bf Limiting cases of Anomalous Hall conductivity and comparison of contributions}
\vskip .2cm
\noindent
In the limit $N_0V_o >> 1$, the results simplify considerably.  The sum of the skew scattering and side jump contributions is then given by

\begin{equation}
\sigma_{xy}^{ss} + \sigma_{xy}^{sj}= e^2\; \sum_{\sigma}\;\tau_{\sigma\sigma}^z\Big[ \frac{1}{2}
\frac{1}{g_{\sigma}\sqrt{w_\sigma}} \Big(\frac{n_\sigma}{n_{imp}}\Big)
 +
\frac{6}{\pi} \frac{1}{g_\sigma(1 + g_\sigma^2/2)w_\sigma^2}\Big]
\label{eq:52}
\end{equation}

\noindent
It is seen that in this case the skew scattering term dominates even in the limit of large impurity concentration.
\vskip .2cm
\noindent
In the case of strong potential scattering, but weak spin-orbit interaction, $w >> 1$, but $u << 1$, we find
\begin{equation}
\sigma_{xy}^{ss} + \sigma_{xy}^{sj} = e^2 \;\sum_{\sigma} \; \tau_{\sigma\sigma}^z
\Big[  \frac{1}{2} \sqrt{u_{\sigma}} \Big(\frac{n_\sigma}{n_{imp}}\Big) + \frac{1}{2\pi}\frac{g_\sigma}{w_\sigma}\Big]\;.
\label{eq:53}
\end{equation}
\vskip .2cm
\noindent
By comparison, the clean limit Hall conductivity $\sigma_{xy}^{c}$ depends only on the spin-orbit coupling $g_{\sigma}$ (see Eq. (19)) or in the case of the Berry phase contribution, on the Berry magnetic field $\mathbf{\Omega}$
(see Eq. (25)).  The former is not dependent on impurity scattering, as well as the latter.  One observes that the signs of $\sigma_{xy}^{c}$ and $\sigma_{xy}^{sj}$ are the same.  In the weak scattering limit $\sigma_{xy}^{sj}$ is a factor of two larger than
$\sigma_{xy}^{c}$, but with increasing scattering strength it drops to values much less than
$\sigma_{xy}^{c}$.  The sign and magnitude of $\sigma_{xy}^B$ have not been calculated for the model under consideration here.
\vskip .2cm
\noindent
Within the model considered here the skew scattering contribution will dominte all other contributions in the limit $n_{imp}\rightarrow 0$.  In a more refined model, however, assuming scattering centers with weak (or no) spin orbit interaction of density $n_n$ in addition to the skew scattering centers with density $n_{imp}$, the mean free path in the limit $n_{imp} \rightarrow 0$ will be limited by the normal scattering processes, and consequently $\sigma_{xy}^{ss} \propto n_{imp}/n_n^2$ and $\sigma_{xy}^{sj}\propto n_{imp}/n_n$ will tend to zero for $n_{imp}\rightarrow 0$.  In this case the clean limit contributions $\sigma_{xy}^c$ and $\sigma_{xy}^B$ will survive.
\vskip .5cm
\section{\bf Conclusion}
\vskip .2cm
\noindent
The anomalous Hall effect is a surprisingly rich phenomenon with many interesting facets.  It requires broken time reversal symmetry as realized in magnetically ordered states and, as far as the symmetry breaking occurs in spin space, it requires sufficiently strong spin-orbit coupling.  The quantum nature of electron scattering by any impurity potential including spin-orbit coupling leads to a right-left asymmetry of the average scattering probability in third order of the potential (``skew scattering'') and to an extra contribution to the velocity operator (``side-jump'' effect).  Both contribute to the Hall conductivity in zero (or low) magnetic field.  In addition, an extra term in the velocity arises directly from the applied electric field, for a uniform system as well as in the periodic potential of the crystal.  The latter contribution has been shown to be a consequence of a Berry phase associated with the motion of Bloch electrons in momentum space.  The extra term in the velocity proportional to the modulus of the applied electric field and directed perpendicular to it has a finite equilibrium expectation value, yielding a Hall current for the clean system.  A systematic experimental investigation showing the existence of all these contributions in a controlled fashion does not exist yet.  Further experimental work on well-characterized systems as a function of magnetization, impurity concentration, for different strengths of spin-orbit interaction is necessary.  Theoretical models for these systems need to be refined, using band structure calculations, to model impurity scattering and spin-orbit interaction in a more realistic way.

\section{\bf acknowledgement}
\vskip .2cm
\noindent
We thank Art Hebard for stimulating our
interest in the anomalous Hall effect and for useful discussions.
KAM thanks U. Karlsruhe for support and hospitality during his
visit.  PW acknowledges partial support through a Max-Planck
Research Award.  This project has also been supported by the DFG
Center for Functional Nanostructures under subproject B2.9.

\end{document}